\documentclass[conference]{IEEEtran}
\IEEEoverridecommandlockouts
\usepackage{cite}
\usepackage{amsmath,amssymb,amsfonts}
\usepackage{algorithmic}
\usepackage{graphicx}
\usepackage{textcomp}
\usepackage{xcolor}

\usepackage{amsthm}
\usepackage{algorithm}
\usepackage{mathrsfs}
\usepackage{multirow}
\usepackage{subfigure} 
\usepackage{amsfonts,amssymb}

\newtheorem{theorem}{Theorem}

\def\BibTeX{{\rm B\kern-.05em{\sc i\kern-.025em b}\kern-.08em
    T\kern-.1667em\lower.7ex\hbox{E}\kern-.125emX}}
\begin{document}

\title{Collaborative Adversarial Learning for Relational Learning on Multiple Bipartite Graphs\\
}

\author{\IEEEauthorblockN{Jingchao Su\textsuperscript{1}, Xu Chen\textsuperscript{1}, Ya Zhang\textsuperscript{*1}, Siheng Chen\textsuperscript{2}, Dan Lv\textsuperscript{3}, and Chenyang Li\textsuperscript{1}}
\IEEEauthorblockA{\textsuperscript{1} \textit{Cooperative Medianet Innovation Center, Shanghai Jiao Tong University, Shanghai, China}\\
\textsuperscript{2} \textit{Mitsubishi Electric Research Laboratories, Cambridge, MA, USA}\\
\textsuperscript{3} \textit{StataCorp LLC, College Station, TX, USA}\\
\{sujingchao, xuchen2016, ya\_zhang\}@sjtu.edu.cn, sihengc@andrew.cmu.edu, dlv@stata.com, lichenyanglh@sjtu.edu.cn
}
}

\maketitle

\begin{abstract}
Relational learning aims to make relation inference by exploiting the correlations among different types of entities. Exploring relational learning on multiple bipartite graphs has been receiving attention because of its popular applications such as recommendations. 
How to make efficient relation inference with few observed links is the main problem on multiple bipartite graphs. 
Most existing approaches attempt to solve the sparsity problem via learning shared representations to integrate knowledge from multi-source data for shared entities. However, they merely model the correlations from one aspect (e.g. distribution, representation), and cannot impose sufficient constraints on different relations of the shared entities.
One effective way of modeling the multi-domain data is to learn the joint distribution of the shared entities across domains.
In this paper, we propose Collaborative Adversarial Learning (CAL) that explicitly models the joint distribution of the shared entities across multiple bipartite graphs. The objective of CAL is formulated from a variational lower bound that maximizes the joint log-likelihoods of the observations. 
In particular, CAL consists of distribution-level and feature-level alignments for knowledge from multiple bipartite graphs. The two-level alignment acts as two different constraints on different relations of the shared entities and facilitates better knowledge transfer for relational learning on multiple bipartite graphs.
Extensive experiments on two real-world datasets have shown that the proposed model outperforms the existing methods.


\end{abstract}


\section{Introduction}
\begin{figure}[tbp]
\centering
\includegraphics[width=0.25\textwidth]{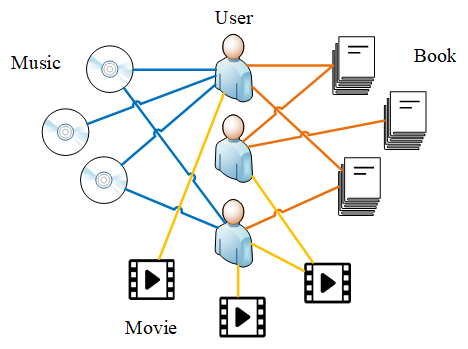} 
\caption{An example of multiple bipartite graphs in recommendation, where user is shared entity type between multiple bipartite graphs. }
\label{instance}
\end{figure}

In the age of abundant yet fragmented information on the Internet, 
it is challenging to explore the relationship between entities from heterogeneous data sources.
As relational learning is an effective technique to learn representations of multi-source data \cite{koller2007introduction},
it is widely applied in various data mining tasks such as recommender systems\cite{singh2008relational}, association prediction\cite{mariappan2019deep}.
In this work, we focus on the scenario with multiple bipartite graphs, where one type of entities is shared between domains while other types have relations with the shared one. The goal is to conduct link prediction between the shared entities and other entities.
A typical example is cross-domain recommendations as Figure \ref{instance} shows, where the domain-shared entities are users and domain-specific entities are music, books and movies. Since a user's limited interaction with items in a certain domain may lead to unsatisfactory recommendation, leveraging his/her information in other domains via relational learning can boost the recommendation performance. 

Relational learning on multiple bipartite graphs can be generally regarded as correlation exploration and knowledge transfer of the shared entities between multiple domains.
The existing methods of integrating the knowledge from different domains mainly fall into the following two categories.
1) \emph{cross-domain representation alignment}, which aims at capturing the relationship across domains through representation alignment \cite{yuan2019darec,frome2013devise,hotelling1992relations,feng2014cross}.
2) \emph{cross-domain representation fusion}, which fuses the features from different domains into a single representation \cite{ngiam2011multimodal,singh2008relational,elkahky2015multi,hu2018conet}. 

In general, the mentioned methods attempt to learn joint representations that are shared in latent space across domains. Despite the promising performance of previous methods, they merely enhance cross-domain correlations from one aspect (e.g. distribution, representation), which does not impose sufficient constraints on different relations of the shared entities. For instance, DARec \cite{yuan2019darec} employs the domain adaptation technique \cite{ganin2016domain} to learn domain-invariant representation via distribution matching. However, distribution matching emphasizes the consistent between the whole, which is inadequate for individual prediction tasks. CMF \cite{singh2008relational} factors relation matrices in all domains and directly shares the representations of shared entities. However, capturing knowledge from different domains with only one representation, the predictive capacity of a single domain may decline, since irrelevant information may be introduced from other domains.
Hence there remains a challenge of how to capture the cross-domain correlation in a holistic way that can effectively improve the prediction performance in each single domain.

Recent work on knowledge transfer \cite{chen2019multivariate,du2018multi} has shown that modeling the joint distribution of entities from different domains facilitates better knowledge transfer because the joint distribution essentially captures the relevance of entities from different domains.
To this end, we propose Collaborative Adversarial Learning (CAL) that explicitly aims this goal. Since the joint distribution is intractable due to its complexity, a variational lower bound is formulated as the optimizing objective for training. In the objective, the latent representations are aligned by matching their parametrized distributions and by enforcing a cross reconstruction. Thus we propose a two-level alignment criterion involving distribution-level and feature-level alignment in agreement with the objective. A CAL model composed of two components---adversarial distribution matching and shared representation matching is built in accordance with the criterion.
Compared with the previous methods, the two-level alignment in our model ensures the integrity of joint distribution matching and a sufficient bidirectional knowledge transfer across domains.

Our major contributions can be summarized as follows: 
\begin{itemize}
\item To explore relational learning on multiple bipartite graphs, we explicitly model the joint distribution of the shared entities. The learning objective is formulated from the variational lower bound that maximizes the joint log-likelihoods.
\item To optimize the variational learning objective, we propose a two-level alignment criterion involving distribution-level and feature-level alignment, and build a model according to the criterion.
\item Experiments are conducted on two real-world datasets to verify the superiority of our model. Further experiments show that our model performs better with three domains than two domains, demonstrating its extensibility for multi-domain tasks.

\end{itemize}

\section{Related Work}
As we regard the relational learning on multiple bipartite graphs as representation learning of the shared entities in different domains, we pay attention to the work of multi-domain representation learning, which is broadly applied in various tasks, e.g. recommender systems, computer visions, cross-media retrieval.
Considering the ways of modeling the correlations across domains, the methods can be mainly divided into cross-domain representation alignment and cross-domain representation fusion \cite{li2018survey}.
\subsection{Cross-domain representation alignment}
Cross-domain representation alignment transforms the original data from different domains into a shared latent space with certain constraints \cite{yuan2019darec,frome2013devise,hotelling1992relations,feng2014cross,andrew2013deep}.
\emph{Correlation-based alignment} aims at maximizing the correlations of representations between domains via CCA \cite{hotelling1992relations}. Further, in order to capture deep non-linear associations between different domains, deep CCA is proposed by using multiple stacked layers of nonlinear mappings \cite{andrew2013deep}.
\emph{Similarity-based alignment} learns a scoring or mapping function to measure the relevance of paired entities across domains \cite{frome2013devise,fu2019deeply,zhu2018deep}. For instance, DeVise \cite{frome2013devise} employs dot-product similarity and hinge rank loss, aiming at producing a higher score for the paired input entities than the unpaired ones.
\emph{Distribution-based alignment} matches the distributions of the representations of shared entities in different domains. A domain classifier is usually employed as a discriminator to distinguish the input from different domains while GRL\cite{ganin2014unsupervised} or GAN\cite{goodfellow2014generative} technique is used to generate representations that deceive the discriminator\cite{yuan2019darec, tzeng2017adversarial}.
\subsection{Cross-domain representation fusion}
Cross-domain representation fusion fuses the features from different domains into a single representation.
\emph{Non-deep-learning-based fusion} learns a probabilistic model over the joint space of the shared latent representations.  
A typical model is Collective Matrix Factorization (CMF) \cite{singh2008relational}, which factors rating matrices in all domains and directly shares the representations of shared entities.
As deep neural networks have been widely applied recently, \emph{deep-learning-based fusion} facilitates stronger inter-domain connections and has shown its superiority for knowledge integration.
\cite{ngiam2011multimodal,elkahky2015multi,lian2017cccfnet,hu2018conet,mariappan2019deep}.
Taking multimodal deep learning \cite{ngiam2011multimodal} for example, it leverages a bimodal deep autoencoder to exploit the concatenated representations from different modality as a shared representation, from which reconstructs both modalities.

\begin{figure}[htb]
\centering
\includegraphics[width=0.30\textwidth]{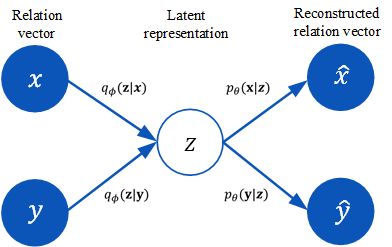} 
\caption{A graphic model of CAL. $\mathbf{x}$ and $\mathbf{y}$ are variables indicating a shared entity's relation in two domains. $\mathbf{z}$ is the shared latent representation of the entity that bridges the gap between different domains.}
\label{graphic}
\end{figure}

\begin{figure*}[htbp]
\centering
\includegraphics[width=0.9\textwidth]{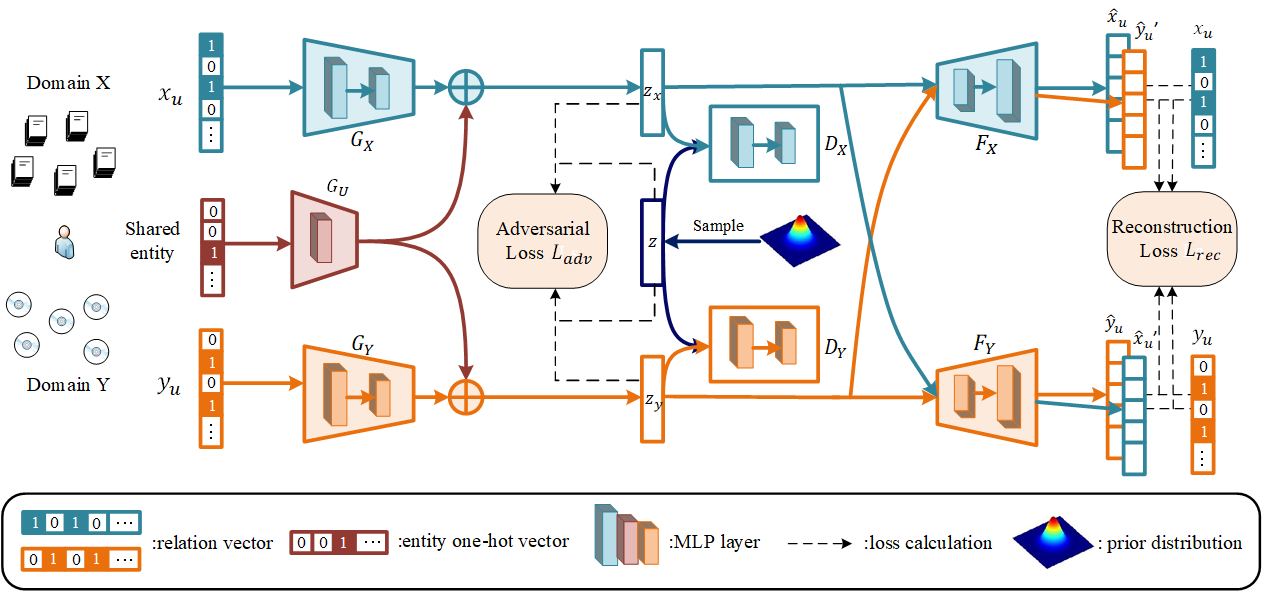} 
\caption{The overall architecture of CAL. Discriminators $D_X$ and $D_Y$ play minimax game with generators $G_X$, $G_Y$ and decoders $F_X$, $F_Y$ to carry out the distribution-level alignment. The stream in blue with $G_X$, $F_X$ and $F_Y$ and the stream in orange with $G_Y$, $F_Y$ and $F_X$ illustrate the feature-level matching. Thus the two-level alignment ensures joint distribution matching. }
\label{model}

\end{figure*}

\section{The proposed model}
In this section, we first describe the relational learning problem on multiple bipartite graphs. 
Then we propose an optimizing objective and formulate a variational lower bound of the objective for training.
Finally we introduce the model framework in accordance with the variational objective, following by the details of each component.

\subsection{Problem Description} 
Consider that there are n+1 types of entities with n bipartite graphs in n domains. $|\mathcal{I^U}|$ entities of type $\mathcal{U}$ are shared in each bipartite graph. Take a two-domain scenario as an example, there are $|\mathcal{I^X}|$ entities in domain $\mathcal{X}$ and $|\mathcal{I^Y}|$ entities in domain $\mathcal{Y}$.
The bipartite graph of domain $\mathcal{X}$ defines the relation between entities in $\mathcal{U}$ and $\mathcal{X}$. It is represented as matrix $\mathbf{X}=\left\{\mathbf{x_{1}},\mathbf{x_{2}},...,\mathbf{x_{|\mathcal{I^U}|}}\right\}$, where $\mathbf{x_{u}}=\left\{x_{u,1},x_{u,2},...,x_{u,|\mathcal{I^X}|}\right\}^\mathrm{T}$ is a vector that describes the relationship between the shared entity $u$ and all the entities in $\mathcal{X}$,
with $x_{u,i}=1$ indicating that $u$ has an observed relationship with entity $x_{i}$ , and $x_{u,i}=0$ indicating not. Similarly, we use the corresponding indications in domain $\mathcal{Y}$. 
Our goal is to exploit the relation among the domains $\mathcal{X}$ and $\mathcal{Y}$ and conduct link prediction by completing matrices $\mathbf{\hat{X}}=\left\{\hat{\mathbf{x_{1}}},\hat{\mathbf{x_{2}}},...,\hat{\mathbf{x_{|\mathcal{I^U}|}}}\right\}$ and $\mathbf{\hat{Y}}=\left\{\hat{\mathbf{y_{1}}},\hat{\mathbf{y_{2}}},...,\hat{\mathbf{y_{|\mathcal{I^U}|}}}\right\}$.


\subsection{The Variational Learning Objective}
\subsubsection{Maximum Joint Likelihood}
To capture the correlations across domains, it is natural to learn to maximize log-likelihood of the joint distribution.
As Figure \ref{graphic} illustrates, $\mathbf{x}$ and $\mathbf{y}$ denote random variables that represent the relation between the shared entity u and the entities in two domains. A shared latent representation $\mathbf{z}$ of entity u is assumed to bridge the gap of the two domains.
The inference and generation nets are presented by conditional distributions $q_{\phi}(\mathbf{z}|\mathbf{x})$, $q_{\phi}(\mathbf{z}|\mathbf{y})$ and $p_{\theta}(\mathbf{x}|\mathbf{z})$, $p_{\theta}(\mathbf{y}|\mathbf{z})$, where $\theta$, $\phi$ denote their parameters. $q_{\phi}(\mathbf{x},\mathbf{y})$ is any joint distribution of random variables ($\mathbf{x},\mathbf{y}$) parametrized by $\phi$.
Given the above description, our goal is to find the parameters so as to maximize the joint log-likelihood $\log p_{\theta}(\mathbf{x},\mathbf{y})$, which can be written as follows:
\begin{equation}
\begin{split}
    \mathbb{E}_{q_{\phi}(\mathbf{x},\mathbf{y})}[\log p_{\theta}(\mathbf{x},\mathbf{y})] 
    =\frac{1}{2}\mathbb{E}_{q_{\phi}(\mathbf{x},\mathbf{y})}[\log p_{\theta}(\mathbf{x}|\mathbf{y}) \\
    + \log p_{\theta}(\mathbf{y})+\log p_{\theta}(\mathbf{y}|\mathbf{x}) + \log p_{\theta}(\mathbf{x})]
\end{split}
\end{equation}
Hence the objective can be decomposed into matching the distribution of marginal likelihoods and conditional likelihoods. 

\subsubsection{The Variational Lower Bound}
For the marginal and conditional distribution matching, the computation requires the marginalization of latent variable $\mathbf{z}$.
However, since the inference of $\mathbf{z}$ is intractable, we resort to the minimization of their variational lower bound. 

\begin{theorem}
\label{thm:lower_bound}
A variational lower bound of Equation (1) is
\begin{eqnarray}
    \mathcal{L} & = & -KL(q_{\phi}(\mathbf{z}|\mathbf{y})||p_{\theta}(\mathbf{z}))-KL(q_{\phi}(\mathbf{z}|\mathbf{x})||p_{\theta}(\mathbf{z})) 
    \\
    && + \mathbb{E}_{q_{\phi}(\mathbf{z}|\mathbf{x})}\left[\log p_{\theta}(\mathbf{x}|\mathbf{z}) + \log p_{\theta}(\mathbf{y}|\mathbf{z})\right] 
    \\
    && +\mathbb{E}_{q_{\phi}(\mathbf{z}|\mathbf{y})}\left[\log p_{\theta}(\mathbf{y}|\mathbf{z}) + \log p_{\theta}(\mathbf{x}|\mathbf{z})\right],
\end{eqnarray}
where $KL()$ indicate Kullback–Leibler divergence.
\end{theorem}
The first two terms in (2) ensure that the distribution of the encoding latent representation should be closed to a prior distribution $p(\mathbf{z})$, and the third term in (3) and the fourth term in (4) indicate that the latent representation encoded from each domain should be decoded to reconstruct the input space in both domains. 
We now sketch the proof.
\begin{proof}
For the conditional distribution matching, the computation requires the marginalization of latent variable z, which is formulated as $p(\mathbf{x}|\mathbf{y})=\int p(\mathbf{x},\mathbf{z}|\mathbf{y})d\mathbf{z}$. However, since the inference of latent variable $\mathbf{z}$ is intractable, we resort to the variational approach, which is achieved by approximating the true posterior distribution $p_{\theta}(\mathbf{z}|\mathbf{x},\mathbf{y})$ by a tractable distribution $q_{\phi}(\mathbf{z}|\mathbf{x},\mathbf{y})$. In addition, we assume $q_{\phi}(\mathbf{z}|\mathbf{x},\mathbf{y})=q_{\phi}(\mathbf{z}|\mathbf{x})=q_{\phi}(\mathbf{z}|\mathbf{y})$, which indicates that the latent representation $\mathbf{z}$ is independent of one domain given another domain.

The optimization of the conditional log-likelihood $p_{\theta}(\mathbf{x}|\mathbf{y})$ can be formulated as
\begin{align}
    &\mathbb{E}_{q_{\phi}(\mathbf{z}|\mathbf{y})}\log p_{\theta}(\mathbf{x}|\mathbf{y}) 
    =\mathbb{E}_{q_{\phi}(\mathbf{z}|\mathbf{y})} \log \frac{p_{\theta}(\mathbf{x},\mathbf{z}|\mathbf{y})}{p_{\theta}(\mathbf{z}|\mathbf{x},\mathbf{y})}\nonumber\\
    = & KL(q_{\phi}(\mathbf{z}|\mathbf{y})||p_{\theta}(\mathbf{z}|\mathbf{x},\mathbf{y}))+\mathbb{E}_{q_{\phi}(\mathbf{z}|\mathbf{y})}\log \frac{p_{\theta}(\mathbf{x},\mathbf{z}|\mathbf{y})}{q_{\phi}(\mathbf{z}|\mathbf{y})} \\
    >&= \mathbb{E}_{q_{\phi}(\mathbf{z}|\mathbf{y})}\log \frac{ p_{\theta}(\mathbf{x},\mathbf{z}|\mathbf{y})}{q_{\phi}(\mathbf{z}|\mathbf{y})}
\end{align}
where KL denotes the KL-divergence and is always non-negative. Equation (6) is the variational lower bound of the conditional log-likelihood, which acts as a surrogate objective function. It can be rewritten as
\vspace{-1mm}
\begin{equation}
\resizebox{.95\linewidth}{!}{$
    \mathcal{L}_{con}=-KL(q_{\phi}(\mathbf{z}|\mathbf{y})||p_{\theta}(\mathbf{z}|\mathbf{y}))+\mathbb{E}_{q_{\phi}(\mathbf{z}|\mathbf{y})}\log p_{\theta}(\mathbf{x}|\mathbf{z},\mathbf{y})
    $}
\end{equation}
The prior of the latent variables z can be relaxed to make the latent variables statistically independent of input variables \cite{kingma2013auto}, denoting as $p(\mathbf{z}|\mathbf{y})=p(\mathbf{z})$. Besides, $p(\mathbf{x}|\mathbf{z},\mathbf{y})=p(\mathbf{x}|\mathbf{z})$ as Figure 1 illustrates. The final objective is
\begin{equation}
    \mathcal{L}_{con}=-KL(q_{\phi}(\mathbf{z}|\mathbf{y})||p_{\theta}(\mathbf{z}))+\mathbb{E}_{q_{\phi}(\mathbf{z}|\mathbf{y})}\log p_{\theta}(\mathbf{x}|\mathbf{z})
\end{equation}
$p(\mathbf{y}|\mathbf{x})$ is optimized in the similar way. 

For the marginal likelihoods, considering the constraint on the latent variable $\mathbf{z}$ which is consistent with the conditional matching, the optimization is the same as the conventional variational autoencoder (VAE)\cite{kingma2013auto}. Its variational lower bound is written as
\begin{align}
    &\mathbb{E}_{q_{\phi}(\mathbf{z}|\mathbf{y})}\log p_{\theta}(\mathbf{y}) \nonumber\\
    = & KL(q_{\phi}(\mathbf{z}|\mathbf{y})||p_{\theta}(\mathbf{z}|\mathbf{y}))+\mathbb{E}_{q_{\phi}(\mathbf{z}|\mathbf{y})}\frac{\log p_{\theta}(\mathbf{y},\mathbf{z})}{q_{\phi}(\mathbf{z}|\mathbf{y})} \\
    \geq & -KL(q_{\phi}(\mathbf{z}|\mathbf{y})||p_{\theta}(\mathbf{z}))+\mathbb{E}_{q_{\phi}(\mathbf{z}|\mathbf{y})}p_{\theta}(\mathbf{y}|\mathbf{z})=\mathcal{L}_{mar}
\end{align}
Incorporating the conditional and marginal distribution matching, the final variational objective is formulated as Theorem 1 presents.
\end{proof}

\subsection{Collaborative Adversarial Learning}
Guided by the the variational lower bound  in Theorem~\ref{thm:lower_bound}, we build Collaborative Adversarial Learning framework; see the the overall framework  in 
Figure~\ref{model}.  This framework can be generally applied to multiple domains. Here we show a model with two domains for simplicity. To optimize the variational objective, a two-level alignment criterion is proposed:




\begin{itemize}
    \item Distribution-level alignment: The whole distributions of shared entities' latent representations, $p(\mathbf{z}^x)$ and $p(\mathbf{z}^y)$, should match a fixed prior distribution, corresponding to Equation (2);
    \item Feature-level alignment: The latent representations of the shared entities $\mathbf{z}_u^x$ and $\mathbf{z}_u^y$ should share the same latent space, corresponding to Equations (3) and (4).
\end{itemize}

Under these conditions, our model mainly consists of two components: 
adversarial distribution matching and shared representation matching. 

\subsubsection{Adversarial Distribution Matching}
According to the distribution-level alignment, the distributions of shared entities' latent representations $p(\mathbf{z}^x)$ and $p(\mathbf{z}^y)$ should be aligned to a fixed prior distribution $p(\mathbf{z})$.
An intuitive example in the application of cross-domain recommendation is that if statistics shows that users prefer pop culture to classics when choosing a movie, a similar tendency will be reflected in their choice of books or musics.
An important approach to distribution matching is adversarial learning \cite{goodfellow2014generative, makhzani2015adversarial}, which employs a generator and a discriminator to play minimax game, until the distribution of latent representations from different domains is indistinguishable. It encouranges the aggregated posterior $p(\mathbf{z}^x)=\int p(\mathbf{z}^x|\mathbf{x})p(\mathbf{x})d\mathbf{x}$ aligned to $p(\mathbf{z})$. Besides, The mode collapses problems is avoided in our model since the reconstruction operation ensures that the latent representation can reconstruct their own input space.



To be concise, we take the domain $\mathcal{X}$ as an example. The discriminator $D_X$ is a domain classifier that predicts the domain label $c \in \left\{0,1\right\}$ of the input vectors $\mathbf{z}_u^x$, where 1 denotes that the vector is drawn from a prior distribution and 0 denotes that the vector belongs to the encoded latent representations in $\mathcal{X}$.
Meanwhile, the generator $G_X$ along with decoder $F_X$ are trying to generate latent representation $\mathbf{z}_u^x$ that cannot be distinguished by the discriminator $D_X$. Therefore, the adversarial training procedure is to minimize domain classifier loss $\mathcal{L}_{adv}$ by optimizing $\theta_D$, and to maximize the same loss by optimizing $\theta_G$,$\theta_F$. $\theta_{G}$,$\theta_{F}$,$\theta_{F}$ are the parameters of the generators, decoders and discriminators respectively. The minimax objective is as follows.

\begin{align}
    &\mathop{\min}_{\theta_{G},\theta_{F}}\mathop{\max}_{\theta_{D}}\mathcal {L}_{adv}(\theta_{G},\theta_{F},\theta_{D})
    =\sum_{u=1}^{|\mathcal{U}|}\left[\log D_X(\mathbf{z}) + \log D_Y(\mathbf{z})\right. \nonumber \\
    & +\left.\log (1-D_X(\mathbf{z}_u^x))  +\log (1-D_Y(\mathbf{z}_u^y))\right]
\end{align}
where $\mathbf{z}$ is randomly drawn from the prior distribution.

\subsubsection{Shared representation Matching}
Under the condition of the feature-level alignment, the latent representation $\mathbf{z}_u^x$/$\mathbf{z}_u^y$ of a shared entity u encoded from each domain is supposed to reconstruct the relation vectors $\mathbf{x}_u$ and $\mathbf{y}_u$ in both domains. 
The condition indicates that given a shared entity's relation in a certain domain, its potential relation in both the current domain and the other domains can be inferenced.
Accordingly, we divide the reconstruction into two parts: self reconstruction and cross reconstruction.


Inspired by Collaborative Denoising AutoEncoder (CDAE) \cite{Yao2016Collaborative}, we corrupt $\tilde{\mathbf{x}}_u$ and $\tilde{\mathbf{y}}_u$ by dropping out the non-zero values of $\mathbf{x}_u$ and $\mathbf{y}_u$ independently with probability q. 
We then feed $\tilde{\mathbf{x}}_u$ into the encoder $G_X$. The shared entity's latent representation $\mathbf{z}_u^x$ is learned by adding a shared-entity embedding $\mathbf{v}_u$ to the output of the encoder $G_X$. And $\mathbf{z}_u^{y}$ is obtained accordingly. $G_X$, $G_Y$ are multi-layer perceptrons. The process is formulated as 
\begin{equation}
\setlength{\abovedisplayskip}{3pt}
\setlength{\belowdisplayskip}{3pt}
\begin{split}
    \mathbf{z}_u^x=G_X(\tilde{\mathbf{x}}_u) + \mathbf{v}_u, \quad
    \mathbf{z}_u^y=G_Y(\tilde{\mathbf{y}}_u) + \mathbf{v}_u.
\end{split}
\end{equation}
The latent representation $\mathbf{z}_u^x$ and $\mathbf{z}_u^y$ are then decoded to reconstruct its relation vectors from both the original domain and the other domain. $F_X$, $F_Y$ are multi-layer perceptrons. The reconstruction loss can be formulated as:
\begin{align}
     \mathcal{L}_{rec} =&\mathcal{L}_{rec}^{x}+\mathcal{L}_{rec}^{y}+\mathcal{L}_{crs}^{xy}+\mathcal{L}_{crs}^{yx} \nonumber\\
     = &\sum_{i=1}^{|\mathcal{U}|}\left[l_{bce}(\mathbf{x}_i,F_X(\mathbf{z}_i^x))+ l_{bce}(\mathbf{y}_i,F_Y(\mathbf{z}_i^y))\right. \nonumber \\
      +& \left. l_{bce}(\mathbf{y}_i,F_Y(\mathbf{z}_i^x))+ l_{bce}(\mathbf{x}_i,F_X(\mathbf{z}_i^y)) \right]
\end{align}
where $l_{bce}$ is the binary cross entropy:
\begin{align}
\resizebox{.95\linewidth}{!}{$
    l_{bce}(\mathbf{a}_u, \hat{\mathbf{a}}_u)=\sum_{i=1}^{N}\left[-a_{u,i}\log \hat{a}_{u,i} - (1-a_{u,i})\log (1- \hat{a}_{u,i})\right]
$}
\end{align}

\subsubsection{Full Objective}
With all the components, the full objective of our model is:
\begin{equation}
\begin{split}
   \mathcal{L}(\theta_{G},\theta_{F},\theta_{})&=\mathcal{L}_{adv}+\mathcal{L}_{rec}
\end{split}
\end{equation}
We aim to solve:
\begin{align}
    (\theta_{G}^*,\theta_{F}^*)
    =\mathop{\arg}\mathop{\min}_{\theta_{G},\theta_{F}}\mathop{\max}_{,\theta_{D}}\mathcal{L}(\theta_{G},\theta_{F},\theta_{D})
\end{align}
where $\theta_{G}^*$ and $\theta_{F}^*$ are optimal parameters of the generators and decoders.

\subsection{Implementation Details}
An illustration of the implemented model is shown in Figure \ref{model}, where $u$ is a shared entity between domain $\mathcal{X}$ and $\mathcal{Y}$. The relation vectors of u, $\mathbf{x_u}$ and $\mathbf{y_u}$ are fed into the the generators $G_X$ and $G_Y$ respectively. 
The latent representation $\mathbf{z_x}$/$\mathbf{z_y}$ is obtained by adding the output from $G_X$/$G_Y$ and the shared-entity embedding from $G_U$.
$\mathbf{z_x}$/$\mathbf{z_y}$ and a randomly sampled $\mathbf{z}$ from a prior distribution are further fed into the discriminators $D_X$/$D_Y$. The discriminators try to distinguish the latent representations $\mathbf{z_x}$/$\mathbf{z_y}$ from $\mathbf{z}$ , while the generators try to generate good representations similar with $\mathbf{z}$ to deceive the discriminators. Besides, the latent representation from each single domain is fed into both decoders $F_X$ and $F_Y$, reconstructing $\mathbf{x_u}$ and $\mathbf{y_u}$ simultaneously. Note that the architecture of generators, decoders and discriminators are implemented by the two-layer perceptrons, and $G_U$ is a learnable embedding lookup table. 

To train our model, we optimize Equation (16) by alternatively updating the parameters of generators \& decoders and discriminators in mini-batch. Stochastic gradient descent (SGD) is adopted to optimize discriminators while ADAM is used to optimize generators. In the testing stage,  the relation in domain $\mathcal{X}$ is predicted by computing $\hat{\mathbf{x}}_u=F_X^*(G_X^*(\mathbf{x}_u))$. And $\hat{\mathbf{y}}_u$ is obtained accordingly.

\begin{table}[t]
\centering
\setlength{\abovecaptionskip}{0pt}%
\setlength{\belowcaptionskip}{5pt}%
\caption{Datasets and Statistics.}\smallskip
\resizebox{\columnwidth}{!}{ 
\begin{tabular}{c|c|c|c|c|c}
\hline
\hline
Dataset & Users & Domain & Items & Interactions & Density \\
\hline
\hline
\multirow{2}*{Douban} & \multirow{2}*{16930} & Movie & 24323 & 2505980 & 0.609\% \\
\cline{3-6}
 & & Book & 40002 & 804818 & 0.119\% \\
\hline
\multirow{2}*{Amazon} & \multirow{2}*{35530} & Movie & 27797 & 724189 & 0.073\% \\
\cline{3-6}
 & & Book & 49697 & 733711 & 0.042\% \\
\hline
\hline
\end{tabular}
 }
\label{table1}

\end{table}  

\section{Experiments}
In this section, we evaluate CAL model on the task of cross-domain recommendation. The users can be viewed as the shared entities and different types of items can be viewed as entities in different domains.
We validate the superiority of our model by comparing it with several state-of-art methods. Sparsity experiments and embedding size experiments are conducted to prove the robustness of our model in handling various situations. Further experiments on multiple domains are also carried out to verify the generality of our model in multi-domain tasks.


\begin{table*}[t]
\centering
\setlength{\abovecaptionskip}{0pt}%
\setlength{\belowcaptionskip}{5pt}%
\caption{Comparasion with baselines on Douban and Amazon. The ranked list is cut off at topN=10.}\smallskip
\resizebox{0.95\textwidth}{!}{ 
\begin{tabular}{c|c|c|c c c c|c c c c c c }
\hline
\hline
Dataset & Domain & Metric & PMF & MLP & CDAE & AAE & CMF & AAE+ & AAE++  & Conet & DARec & CAL \\
\hline
\hline
\multirow{6}*{Douban} & \multirow{3}*{Movies} & hr & 
 0.8711 & 0.8746 & 0.8955 & 0.8961 & 0.8963 & 0.8853& 0.9005 & 0.8944 & 0.9003 & \textbf{0.9028}  \\
\cline{3-13}
 & & ndcg & 
 0.6281 & 0.6286 & 0.7038 & 0.6919 & 0.6699 & 0.6683 & 0.7037 & 0.6725 & 0.7004 & \textbf{0.7077} \\
\cline{3-13} 
 & & mrr &
 0.5509 & 0.5505 & 0.6426 & 0.6269 & 0.5675 & 0.5991 & 0.6408 & 0.6112 & 0.6365 & \textbf{0.6454}  \\
\cline{2-13} 
 & \multirow{3}*{Books} & hr & 
 0.7579 & 0.7325 & 0.7908 & 0.7829 & 0.8017 & 0.7609 & 0.7906 & 0.7644 & 0.7877 & \textbf{0.8033}     \\
 \cline{3-13}
 & & ndcg & 
 0.5573 & 0.5250 & 0.6199 & 0.6054 & 0.6034 & 0.5776 & 0.6223 & 0.5651 & 0.6130 & \textbf{0.6300}  \\
\cline{3-13} 
 & & mrr & 
 0.4938 & 0.4597 & 0.5660 & 0.5469 & 0.5406 & 0.5197 & 0.5694 & 0.5234 & 0.5577 & \textbf{0.5753}    \\
 \hline
 \hline
 \multirow{6}*{Amazon} & \multirow{3}*{Movies} & hr &
 0.6068 & 0.5349 & 0.6349 & 0.6050 & 0.6127 & 0.6006 & 0.6331 & 0.5771 & 0.6411& \textbf{0.6541}\\
\cline{3-13}
 & & ndcg &
 0.3792 & 0.3214 & 0.4222 & 0.3918 & 0.3850 & 0.3852 & 0.4193 & 0.3210 & 0.4202& \textbf{0.4338}\\
\cline{3-13} 
 & & mrr &
 0.3089 & 0.2558 & 0.3563 & 0.3259 & 0.3147 & 0.3187 & 0.3532 & 0.3520 & 0.3519& \textbf{0.3654}\\
\cline{2-13} 
 & \multirow{3}*{Books} & hr &
 0.4983 & 0.4103 & 0.5410 & 0.5191 & 0.5409 & 0.5298 & 0.5297 & 0.4668 & 0.5350 & \textbf{0.5995}\\
 \cline{3-13}
 & & ndcg &
 0.3166 & 0.2419 & 0.3482 & 0.3310 & 0.3411 & 0.3332 & 0.3431 & 0.2720 & 0.3454 & \textbf{0.3905}\\
\cline{3-13} 
 & & mrr &
 0.2606 & 0.1904 & 0.2889 & 0.2731 & 0.2795 & 0.2727 & 0.2856 & 0.2787 & 0.2871 & \textbf{0.3260}\\
 \hline
 \hline
\end{tabular}
}
\label{table2}

\end{table*}


\subsection{Dataset}
\begin{itemize}
    \item \textbf{Douban} is a Chinese social networking service website which contains users' rating records of movies, books, etc. We select users that rate both movies and books and convert the rated items to positive samples for each user.
    \item \textbf{Amazon} is an E-commerce platform selling multiple categories of product. We choose the categories of Movies and Books since they contain rich rating data among all categories. For multi-domain experiments, the users who have records in all the categories of Movies , Books, Music are chosen.
    The positive samples are obtained in the same way as Douban dataset. 
\end{itemize}




\subsection{Baselines}
We test our method with the following baselines of both single-domain recommendation and cross-domain recommendation.
\subsubsection{Methods for single-domain recommendation}

\begin{figure*}[htbp]
\centering
 
\subfigure[Douban Movie HR@10]{
\begin{minipage}[t]{0.25\textwidth}
\centering
\includegraphics[width=\textwidth]{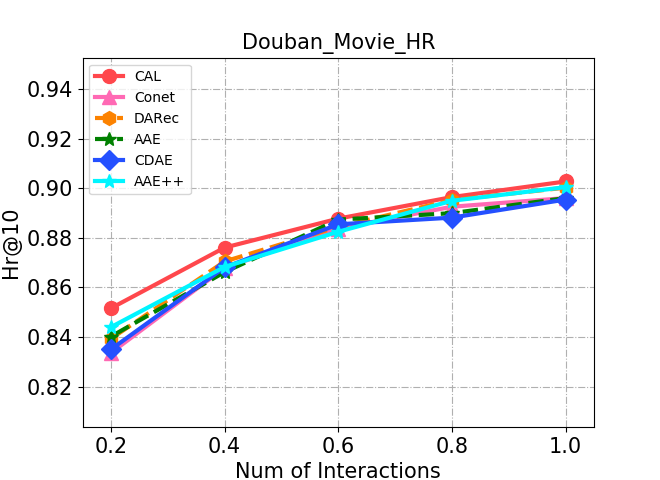}
\end{minipage}%
}%
\subfigure[Douban Movie NDCG@10]{
\begin{minipage}[t]{0.25\textwidth}
\centering
\includegraphics[width=\textwidth]{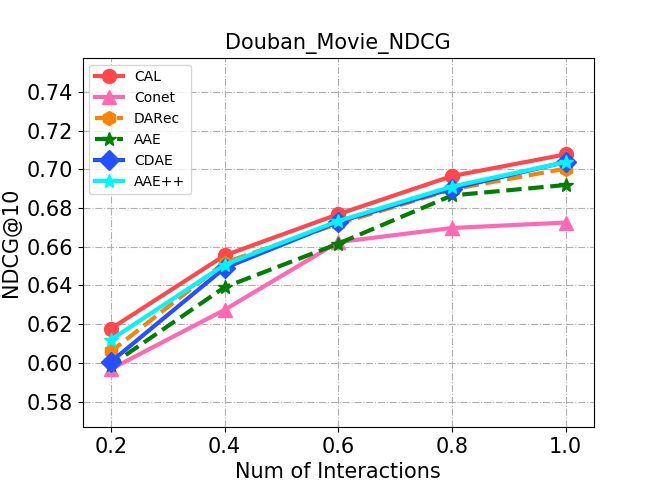}
\end{minipage}%
}%
\subfigure[Douban Books HR@10]{
\begin{minipage}[t]{0.25\textwidth}
\centering
\includegraphics[width=\textwidth]{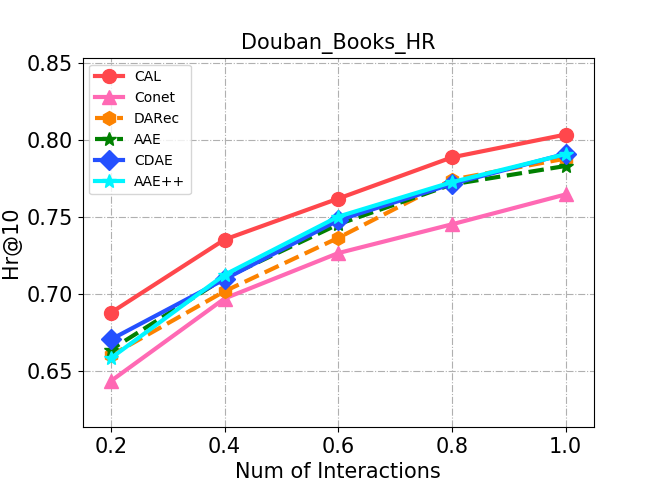}
\end{minipage}%
}%
\subfigure[Douban Books NDCG@10]{
\begin{minipage}[t]{0.25\textwidth}
\centering
\includegraphics[width=\textwidth]{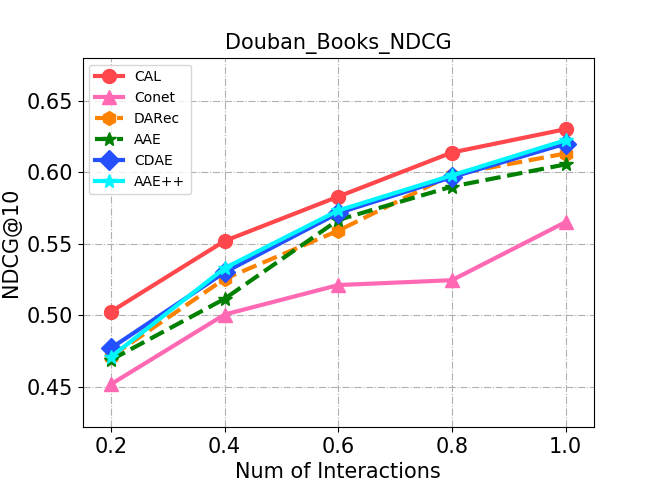}
\end{minipage}%
}%

\subfigure[Amazon Movie HR@10]{
\begin{minipage}[t]{0.25\textwidth}
\centering
\includegraphics[width=\textwidth]{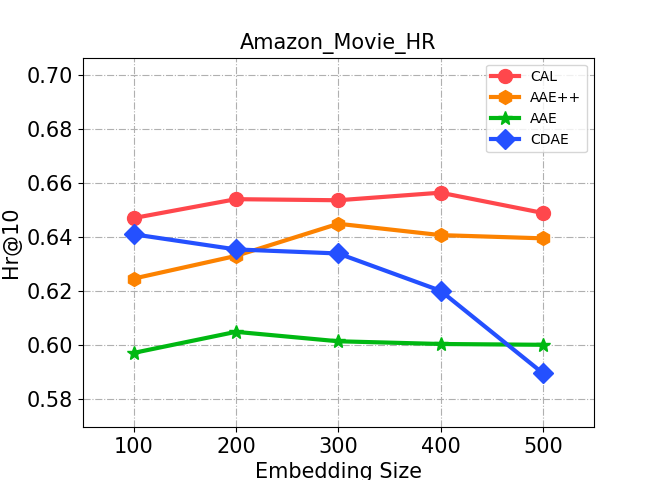}
\end{minipage}
}%
\subfigure[Amazon Movie NDCG@10]{
\begin{minipage}[t]{0.25\textwidth}
\centering
\includegraphics[width=\textwidth]{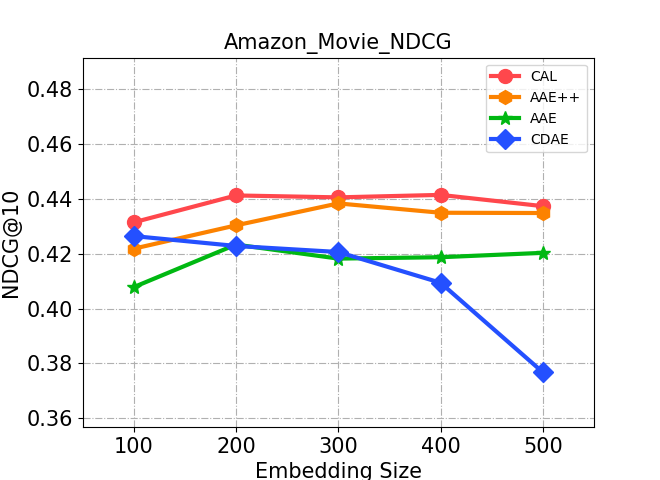}
\end{minipage}
}%
\subfigure[Amazon Books HR@10]{
\begin{minipage}[t]{0.25\textwidth}
\centering
\includegraphics[width=\textwidth]{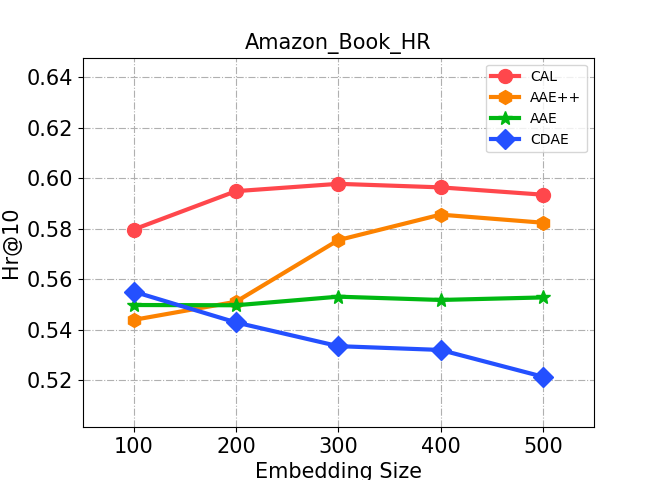}
\end{minipage}
}%
\subfigure[Amazon Books NDCG@10]{
\begin{minipage}[t]{0.25\textwidth}
\centering
\includegraphics[width=\textwidth]{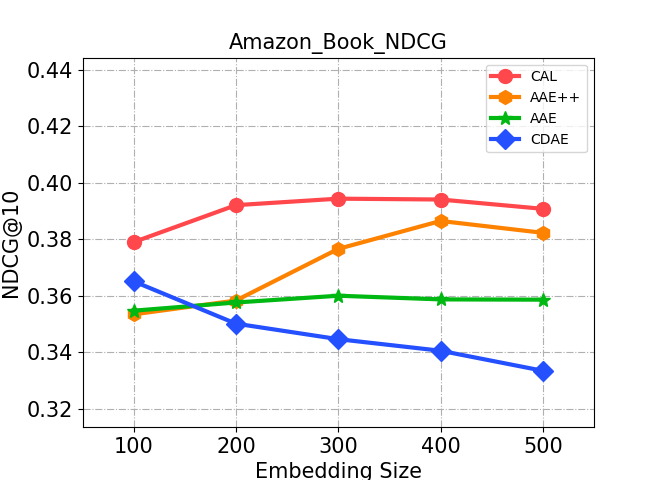}
\end{minipage}
}%
 
\centering
\caption{Sparsity analysis when the number of user-item interaction ranges from 20\% to 100\% on Douban and Amazon.}
\label{sparsity}
\end{figure*}
\begin{itemize}
    \item \textbf{PMF} \cite{mnih2008probabilistic} is a widely-used model for matrix completion. The missing relations are predicted via the inner product of the interacted entities' representations.
    \item \textbf{MLP} \cite{he2017neural} is a basic deep learning approach that models the relation of interacted entities via feeding the concatenation of their representations into multilayer neural networks.
    \item \textbf{CDAE} \cite{wu2016collaborative} is an denoising extension of the classical Auto-Encoder and is designed for top-N recommendation. 
    \item \textbf{AAE} \cite{makhzani2015adversarial} uses GAN to match the aggregated posterior of the latent vector with a prior distribution. We modify AAE for our top-N recommendation task by replacing its Auto-Encoder with CDAE. 
\end{itemize}
\subsubsection{Methods for cross-domain recommendation}
\begin{itemize}
    \item \textbf{CMF} \cite{singh2008relational} simultaneously factors matrices in different domains and shares representations among shared entities for handling information from multiple relations.
    \item \textbf{Conet} \cite{hu2018conet} is an approach based on MLP. It enables dual knowledge transfer across domains via a cross switch network between the adjacent layers in different domains.
    \item \textbf{DARec} \cite{yuan2019darec} is the state-of-the-art method. It leverages domain adversarial neural networks \cite{ganin2014unsupervised} to extract the shared representations from two domains. We replace its encoder with CDAE to apply for the top-N recommendation scenario.
    \item \textbf{AAE+} takes the union of entities in all domains and runs a AAE model.
    \item \textbf{AAE++} combines two AAE models \cite{makhzani2015adversarial} and shares their prior distribution. It is a degraded model of CAL without cross reconstruction loss.
\end{itemize}
\subsection{Experimental Settings}
\subsubsection{Evaluation Protocol}
For top-N recommendation task, we adopt the \textit{leave-one-out} evaluation, which is widely used in the literature of recommendation \cite{rendle2009bpr,he2017neural}. For each user, one interacted item is reserved as a positive valid sample and 99 non-interacted items are randomly chosen as negative valid samples. We evaluate the performance by ranking the positive valid sample among the 100 items and compute its hit ratio (HR), normalized discounted cumulative gain (NDCG) and mean reciprocal rank (MRR). 

\subsubsection{Experimental Implementation}
For PMF, MLP, CDAE, AAE, CMF and Conet, we use the code released by the author. AAE+ and AAE++ is implemented based on AAE. DARec is coded referring to the paper. For PMF, MLP, CMF and Conet which require feeding user-item pairs, we utilize a negative sampling approach, in which 5 negative instances are sampled per positive instance.

Our model is implemented with Pytorch. The optimizer for discriminators is SGD and the optimizer for the rest of the networks is ADAM. The parameters are updated in mini-batch and the learning rate is 0.001. The prior distribution here is set as Standard Gaussian $\mathcal{N}(0,1)$, which is a common setting in recent adversarial-based method \cite{goodfellow2014generative,makhzani2015adversarial}. For CDAE, AAE, AAE+, AAE++ and CAL, the layers of encoders and decoders are 2, and the embedding size is 200. 


\begin{figure*}[htbp]
\centering
\setlength{\abovecaptionskip}{-0mm}%
\setlength{\belowcaptionskip}{-5mm}%
\subfigure[Amazon Movie HR@10]{
\begin{minipage}[t]{0.25\textwidth}
\centering
\includegraphics[width=\textwidth]{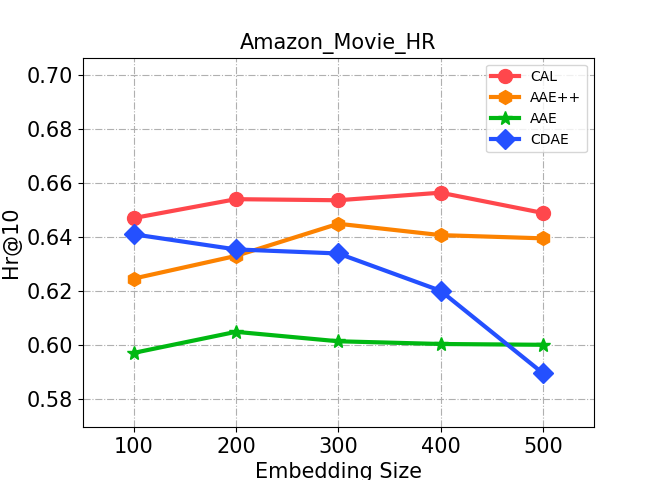}
\end{minipage}
}%
\subfigure[Amazon Movie NDCG@10]{
\begin{minipage}[t]{0.25\textwidth}
\centering
\includegraphics[width=\textwidth]{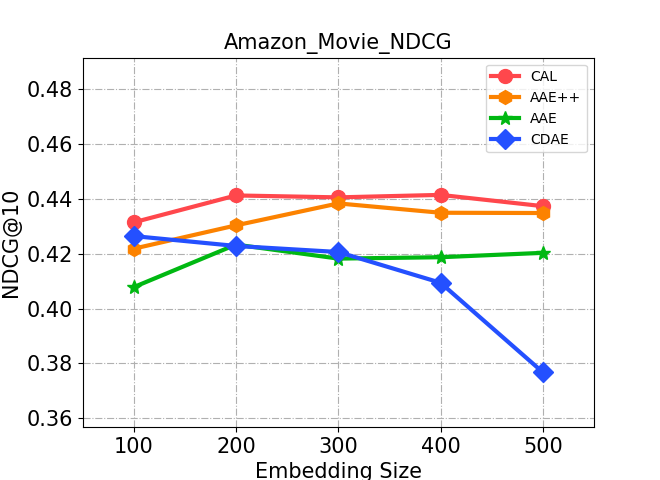}
\end{minipage}
}%
\subfigure[Amazon Books HR@10]{
\begin{minipage}[t]{0.25\textwidth}
\centering
\includegraphics[width=\textwidth]{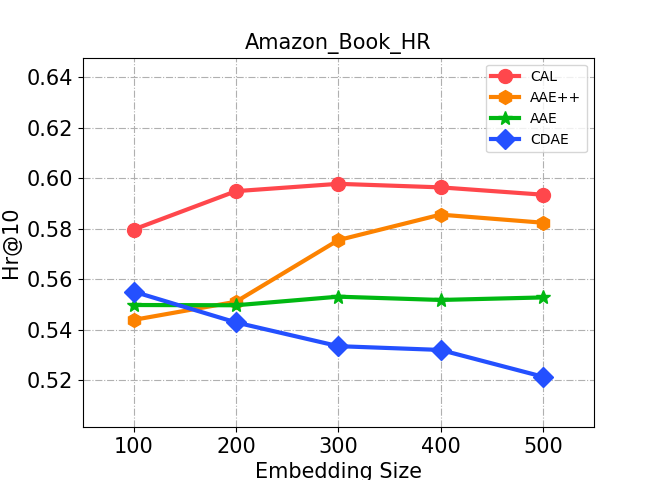}
\end{minipage}
}%
\subfigure[Amazon Books NDCG@10]{
\begin{minipage}[t]{0.25\textwidth}
\centering
\includegraphics[width=\textwidth]{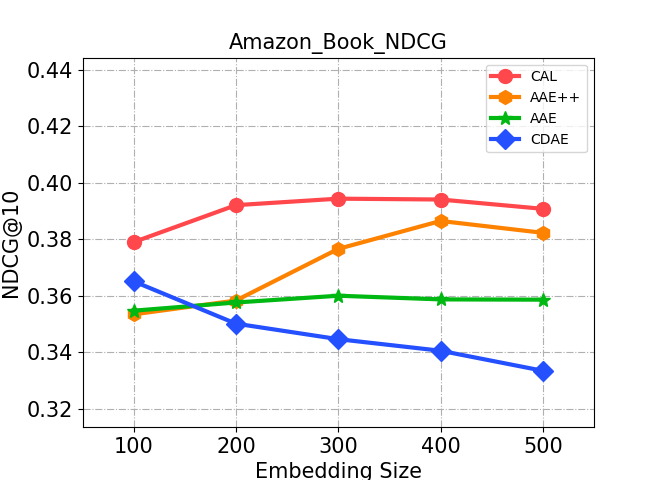}
\end{minipage}
}%
 
\centering
\caption{Embedding size analysis when the embedding size of users' latent representations varies from 100 to 500 on Amazon.}
\label{emb}
\end{figure*}

\vspace{-2mm}
\subsection{Overall Performance}
Table \ref{table2} shows the results of different methods on the two datasets under three ranking metrics, from which we have the following findings:
\begin{itemize}
    \item The cross-domain methods outperform their corresponding single domain approaches that are similar in structure (e.g., PMF \& CMF, MLP \& Conet, AAE \& AAE++). This suggests that relational learning on multiple bipartite graphs facilitates knowledge transfer from relevant domains, which enriches the training data and thus improves the recommendation performance.
    
    \item Comparing deep methods with shallow ones, we find the shallow methods get relatively low NDCG and MRR even their HR do not make much difference in Douban. It indicates that shallow methods may be appropriate for link prediction in dense datasets, but are not as satisfactory in sparse datasets or in ranking tasks.
    
    \item AAE++ gets better performance compared to AAE+, while the main difference is that AAE++ utilizes one network per domain and AAE+ uses one for all domains. This shows that the shared representations of different domains is hard to learn from only one network due to the distinction between domains.
    CAL achieves superior performance than AAE++,
    while the difference is that CAL employs an extra cross reconstruction loss.
    This demonstrates the importance of the two-level alignment of CAL for the adequate transfer of knowledge, as it allows for more holistic joint distribution matching. This also accounts for our better performance than DARec.

    \item CAL outperforms all the single-domain and cross-domain methods in both datasets. In particular, CAL improves over the state-of-the-art methods with a 0.28\%, 1.98\%, 2.03\%, 10.83\%  relative gain of HR@10 in Douban Movie, Douban Book, Amazon Movie and Amazon Book respectively. Generally the gain is more obvious as the data gets sparser. The superior performance of CAL shows its strong ability of relation inference between bipartite graphs, especially in sparse datasets.
\end{itemize}

\subsection{Sparsity Analysis}
To investigate whether CAL can still perform better than other methods under
more sparse conditions, we conduct sparsity experiments by gradually reducing the quantity of the target bipartite graph's links. Specifically, we randomly draw 80\%, 60\%, 40\% and 20\% of the links as training sets. 
Our method is compared with the competitive methods for both single-domain and cross-domain methods respectively, and the results are shown in Figure \ref{sparsity}. 

CAL consistently outperforms other methods on almost all sparsity conditions.
In particular, all the methods do not make much difference in dense Douban Movie dataset. But as the dataset gets sparser and the number of user-item links decreases, our method increasingly outperforms baselines in most of the situations. However, in the extreme sparse case 20\% in Amazon, the gap among these methods is not obvious, because the observed links in target bipartite graph are too few to train a reliable model even the other bipartite graphs have rich links.
 

\begin{table}[tbp]
\centering
\setlength{\abovecaptionskip}{0pt}%
\setlength{\belowcaptionskip}{5pt}%
\caption{Statistics for Three Domains in Amazon.}\smallskip
\resizebox{\columnwidth}{!}{ 
\begin{tabular}{c|c|c|c|c|c}
\hline
\hline
Dataset & Users & Domain & Items & Interactions & Density \\
\hline
\hline
\multirow{3}*{Amazon} & \multirow{3}*{11781} & Movies & 18411 & 359348 &  0.166\% \\
\cline{3-6}
 & & Books & 20648 & 229559 & 0.094\% \\
\cline{3-6}
 & & Music & 13708 & 168504 &  0.104\% \\
\hline
\hline
\end{tabular}
}
\label{table3}
\end{table}

\subsection{Embedding Size Analysis}
The dimension of the latent representations is an important factor for representation learning. We thus investigate the performance of CDAE, AAE, AAE++, and CAL under different users embedding sizes on Amazon dataset. The results are shown in Figure \ref{emb}.

The results show the adaptive abilities of the tested methods to the changes of the embedding size. The performance of CDAE declines rapidly with the increase of the embedding size due to overfitting. AAE is stable but does not perform well. AAE++ suffer from underfitting when the embedding size is below 300 for movie domain and below 400 for book domain. CAL, however, is stable with preeminent performance regardless of the changes, showing its capability of handling the latent representations with different embedding sizes.


\begin{table}[t]
\centering
\small
\setlength{\abovecaptionskip}{1pt}%
\setlength{\belowcaptionskip}{5pt}%
\caption{Three-domains Recommendation in Amazon.}\smallskip
\resizebox{\columnwidth}{!}{ 
\begin{tabular}{c|c|c c c c}
\hline
\hline
\multirow{2}*{Domain} & \multirow{2}*{Metric} & Movie & Book & Music & Three \\
 & & \&Book & \&Music & \&Movie & domains \\
\hline
\hline
\multirow{3}*{Movies} & hr & 
 0.5725  & -  & 0.5791  & \textbf{0.5795}    \\
\cline{2-6}
 & ndcg & 
 0.3566 & - & \textbf{0.3627} & 0.3601  \\
\cline{2-6} & mrr &
 0.2900 & - & \textbf{0.2960} & 0.2927   \\
\cline{1-6}\multirow{3}*{Books} & hr & 
 0.5448 & 0.5302 & - & \textbf{0.5478}    \\
\cline{2-6}
& ndcg & 
 0.3416 & 0.3356 & - & \textbf{0.3456}   \\
\cline{2-6}
 & mrr & 
 0.2792 & 0.2758 & - & 0.\textbf{2834}  \\
 \cline{1-6} 
 \multirow{3}*{Music} & hr & 
 - & 0.5253 & 0.5422 & \textbf{0.5499}    \\
 \cline{2-6}
 & ndcg & 
 - & 0.3281 & 0.3420 & \textbf{0.3441}   \\
\cline{2-6} 
 & mrr & 
 - & 0.2675 & 0.2806 & \textbf{0.2807}  \\
 \hline
 \hline
\end{tabular}
}
\label{table4}
\end{table}

\subsection{Multi-Domain Analysis}
To prove the generality of CAL on more bipartite graphs, we first conduct experiments respectively on each two of the three domains in Amazon dataset. The model is then trained with all three domains. Tables \ref{table3} and \ref{table4} show the statistics of the data and the result of the experiments respectively. 

In most of the cases, the introduction of the bipartite graphs in extra domains can improve the performance. This indicates that although more domains may bring more noise, our model can effectively extract useful knowledge from heterogeneous data, which demonstrating its extensibility on multiple domains.
However, for movie recommendation, the result of music \& movie slightly surpasses that of three domains combination. This is intuitively reasonable because when people watch a movie, they tend to listen to the sound track of this movie, leading to a closer connection between these two bipartite graphs. But it should still be noticed that an increasing number of bipartite graphs may lead to the decline of the performance, for the reason that their shared representations are inclined to capture only the shared knowledge among domains.

\section{Conclusions and future work}
In this paper, we propose CAL to explore relational learning among multiple bipartite graphs. We aim to explicitly model the joint distribution of the shared entities as it intrinsically captures the relevance between bipartite graphs. The learning objective is formulated from the variational lower bound of maximizing the joint log-likelihoods and a two-level alignment criterion involving distribution-level and feature-level is proposed accordingly. Experiments are conducted on two real-world datasets and the results confirm the superiority of CAL compared with state-of-the-art methods.

Despite the superiority of CAL, there are some aspects that could be studied further. For example, to prove the effectiveness and generality of CAL, we make a simple assumption about the prior distribution as a standard Gaussian distribution. However, real-world data tends to have more complex structures that are difficult to capture with such simple distribution. Therefore, we will explore better ways to model the prior distribution for more comprehensive data mining on multiple bipartite graphs.

Translated with www.DeepL.com/Translator (free version)

\bibliographystyle{IEEEtran}
\bibliography{ickg}

\end{document}